\documentstyle[prd,tighten,aps,amssymb,newlfont,epsfig]{revtex}
\begin{document}
\draft
\preprint{
\begin{tabular}{r}
   UWThPh-1999-20
\\ DFTT 17/99
\\ hep-ph/9903454
\end{tabular}
}
\twocolumn[\hsize\textwidth\columnwidth\hsize\csname
@twocolumnfalse\endcsname
\title{Four-Neutrino Mass Spectra and the Super-Kamiokande 
Atmospheric Up--Down Asymmetry}
\author{S.M. Bilenky}
\address{Joint Institute for Nuclear Research, Dubna, Russia, and\\
Institute for Theoretical Physics, University of Vienna,\\
Boltzmanngasse 5, A--1090 Vienna, Austria}
\author{C. Giunti}
\address{INFN, Sezione di Torino, and Dipartimento di Fisica Teorica,
Universit\`a di Torino,\\
Via P. Giuria 1, I--10125 Torino, Italy}
\author{W. Grimus and T. Schwetz}
\address{Institute for Theoretical Physics, University of Vienna,\\
Boltzmanngasse 5, A--1090 Vienna, Austria}
\date{22 March 1999}
\maketitle
\begin{abstract}
In the framework of schemes with mixing of four massive neutrinos,
which can accommodate atmospheric, solar and LSND ranges of 
$\Delta m^2$, we show that, 
in the whole region of $\Delta m^2_\mathrm{LSND}$ allowed by LSND,
the Super-Kamiokande up--down asymmetry
excludes all mass spectra with a group of three close
neutrino masses separated from the fourth mass by the LSND gap of order
1~eV. Only two schemes with mass spectra in which two pairs of close masses 
are separated by the LSND gap can describe the Super-Kamiokande
up--down asymmetry and all other existing neutrino oscillation data. 
\end{abstract}
\pacs{UWThPh-1999-20, DFTT 17/99, hep-ph/9903454}
]

The observation of a significant up--down asymmetry of atmospheric high-energy
$\stackrel{\scriptscriptstyle (-)}{\nu}_{\hskip-3pt \mu}$-induced events 
in the Super-Kamiokande experiment \cite{SK-atm-98}
is considered as the first model-independent evidence in 
favor of neutrino oscillations.
Such indications were also obtained in other atmospheric
neutrino experiments:
Kamiokande \cite{Kam-atm-94},
IMB \cite{Bec95},
Soudan-2 \cite{Soudan2}
and
MACRO \cite{MACRO-98}.
In addition, evidence in favor of neutrino masses and mixing is
provided by all solar neutrino experiments:
Homestake \cite{Homestake98},
Kamiokande \cite{Kam-sun-96},
GALLEX \cite{GALLEX96},
SAGE \cite{SAGE96} and
Super-Kamiokande \cite{SK-sun-98-PRL}.
Finally, observation of
$\bar\nu_\mu\to\bar\nu_e$
and
$\nu_\mu\to\nu_e$
oscillations
have been claimed by the LSND collaboration \cite{LSND-4}.
For the explanation of all these data three different scales of neutrino
mass-squared differences are required:
$ \Delta{m}^2_{\mathrm{sun}} \sim 10^{-10} \, \mathrm{eV}^2 $ 
(vacuum oscillations)
or
$ \Delta{m}^2_{\mathrm{sun}} \sim 10^{-5} \, \mathrm{eV}^2 $ (MSW),
$ \Delta{m}^2_{\mathrm{atm}} \sim 10^{-3} \, \mathrm{eV}^2 $,
$ \Delta{m}^2_{\mathrm{LSND}} \sim 1 \, \mathrm{eV}^2 $.
Thus, at least four neutrinos with definite mass are needed to
describe all data.

\begin{figure}[t]
\begin{center}
\setlength{\unitlength}{1.0cm}
\begin{tabular*}{0.99\linewidth}{@{\extracolsep{\fill}}cccccc}
\begin{picture}(1,4) 
\thicklines
\put(0.1,0.2){\vector(0,1){3.8}}
\put(0.0,0.2){\line(1,0){0.2}}
\put(0.4,0.15){\makebox(0,0)[l]{$m_1$}}
\put(0.0,0.4){\line(1,0){0.2}}
\put(0.4,0.45){\makebox(0,0)[l]{$m_2$}}
\put(0.0,0.8){\line(1,0){0.2}}
\put(0.4,0.8){\makebox(0,0)[l]{$m_3$}}
\put(0.0,3.5){\line(1,0){0.2}}
\put(0.4,3.5){\makebox(0,0)[l]{$m_4$}}
\end{picture}
&
\begin{picture}(1,4) 
\thicklines
\put(0.1,0.2){\vector(0,1){3.8}}
\put(0.0,0.2){\line(1,0){0.2}}
\put(0.4,0.2){\makebox(0,0)[l]{$m_1$}}
\put(0.0,0.6){\line(1,0){0.2}}
\put(0.4,0.55){\makebox(0,0)[l]{$m_2$}}
\put(0.0,0.8){\line(1,0){0.2}}
\put(0.4,0.85){\makebox(0,0)[l]{$m_3$}}
\put(0.0,3.5){\line(1,0){0.2}}
\put(0.4,3.5){\makebox(0,0)[l]{$m_4$}}
\end{picture}
&
\begin{picture}(1,4) 
\thicklines
\put(0.1,0.2){\vector(0,1){3.8}}
\put(0.0,0.2){\line(1,0){0.2}}
\put(0.4,0.2){\makebox(0,0)[l]{$m_1$}}
\put(0.0,2.9){\line(1,0){0.2}}
\put(0.4,2.9){\makebox(0,0)[l]{$m_2$}}
\put(0.0,3.3){\line(1,0){0.2}}
\put(0.4,3.25){\makebox(0,0)[l]{$m_3$}}
\put(0.0,3.5){\line(1,0){0.2}}
\put(0.4,3.55){\makebox(0,0)[l]{$m_4$}}
\end{picture}
&
\begin{picture}(1,4) 
\thicklines
\put(0.1,0.2){\vector(0,1){3.8}}
\put(0.0,0.2){\line(1,0){0.2}}
\put(0.4,0.2){\makebox(0,0)[l]{$m_1$}}
\put(0.0,2.9){\line(1,0){0.2}}
\put(0.4,2.85){\makebox(0,0)[l]{$m_2$}}
\put(0.0,3.1){\line(1,0){0.2}}
\put(0.4,3.15){\makebox(0,0)[l]{$m_3$}}
\put(0.0,3.5){\line(1,0){0.2}}
\put(0.4,3.5){\makebox(0,0)[l]{$m_4$}}
\end{picture}
&
\begin{picture}(1,4) 
\thicklines
\put(0.1,0.2){\vector(0,1){3.8}}
\put(0.0,0.2){\line(1,0){0.2}}
\put(0.4,0.2){\makebox(0,0)[l]{$m_1$}}
\put(0.0,0.6){\line(1,0){0.2}}
\put(0.4,0.6){\makebox(0,0)[l]{$m_2$}}
\put(0.0,3.3){\line(1,0){0.2}}
\put(0.4,3.25){\makebox(0,0)[l]{$m_3$}}
\put(0.0,3.5){\line(1,0){0.2}}
\put(0.4,3.55){\makebox(0,0)[l]{$m_4$}}
\end{picture}
&
\begin{picture}(1,4) 
\thicklines
\put(0.1,0.2){\vector(0,1){3.8}}
\put(0.0,0.2){\line(1,0){0.2}}
\put(0.4,0.15){\makebox(0,0)[l]{$m_1$}}
\put(0.0,0.4){\line(1,0){0.2}}
\put(0.4,0.45){\makebox(0,0)[l]{$m_2$}}
\put(0.0,3.1){\line(1,0){0.2}}
\put(0.4,3.1){\makebox(0,0)[l]{$m_3$}}
\put(0.0,3.5){\line(1,0){0.2}}
\put(0.4,3.5){\makebox(0,0)[l]{$m_4$}}
\end{picture}
\\
(I) & (II) & (III) & (IV) & (A) & (B)
\end{tabular*}
\end{center}
\caption{ \label{4spectra}
The six types of neutrino mass spectra that can accommodate 
the solar, atmospheric and LSND scales of $\Delta{m}^2$. The different
distances between the masses on the vertical axes symbolize the
different scales of $\Delta{m}^2$. The spectra (I) -- (IV) define
class 1, whereas class 2 comprises (A) and (B).}
\end{figure}
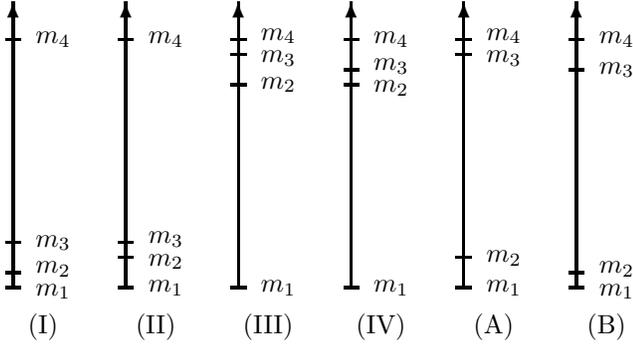

Four-neutrino schemes have been considered in many papers. For early
works see Ref.~\cite{4-early} and for a more comprehensive list of
four-neutrino papers consult, \textit{e.g.}, Ref.~\cite{BGG-review}.
In Refs.~\cite{BGG96,Okada-Yasuda97,BPWW98-PRD58} it was shown that
from the results of all existing experiments,
including short-baseline (SBL)
reactor and accelerator experiments
in which no indications of neutrino oscillations have been found,
information on the four-neutrino mass spectrum can be inferred.
In the case of three different scales of $\Delta m^2$,
there are two different classes of neutrino mass spectra
(see Fig.~\ref{4spectra})
that satisfy the inequalities
$ \Delta{m}^2_{\mathrm{sun}} \ll \Delta{m}^2_{\mathrm{atm}} 
\ll \Delta{m}^2_{\mathrm{LSND}} $.
In the spectra of class 1 there is a group of three close masses
which is separated from the fourth mass by the 
LSND gap of around 1 eV. It contains the spectra (I) -- (IV)
in Fig.~\ref{4spectra}. Note that spectrum (I) corresponds to a mass
hierarchy, spectrum (III) to an inverted mass hierarchy, whereas (II) and
(IV) are non-hierarchical spectra.
In the spectra of class 2 there are two pairs of close masses
which are separated by the LSND gap. The two possible spectra
in this class are denoted by (A) and (B) in Fig.~\ref{4spectra}.

It was shown in Ref.~\cite{BGG96}
that, in the case of the spectra of class 1, from the existing data
one can obtain
constraints on the amplitude of SBL
$\nu_\mu\to\nu_e$ oscillations   
that are not compatible with the results of the LSND experiment
in the allowed region 
$0.2 \: \mathrm{eV}^2 \lesssim \Delta{m}^2_{\mathrm{LSND}}
\lesssim 2 \: \mathrm{eV}^2$
with the exception of the small interval from 0.2 to
0.3 eV$^2$. In Ref.~\cite{BGG96} the double ratio $R$ of
$\mu$-like over $e$-like events has been used as input from atmospheric
neutrino measurements, whereas in the present letter we 
consider what constraints on neutrino mixing
can be inferred from the up--down asymmetry of multi-GeV muon-like events
measured in the Super-Kamiokande experiment \cite{UDA99}, \textit{i.e.},
from 
\begin{equation}\label{A}
A_\mu = \frac{U-D}{U+D} = - 0.311 \pm 0.043 \pm 0.01
\,,
\end{equation}
where $U$ and $D$ denote the number of events in the zenith angle
ranges $-1 < \cos \theta < -0.2$ and $0.2 < \cos \theta < 1$,
respectively. We will show that with this input the conclusion
of Ref.~\cite{BGG96} will be strengthened and that now the 
neutrino mass spectra of class 1
are disfavored for any value of $\Delta{m}^2_{\mathrm{LSND}}$
in the allowed range. In addition, we will also derive a
constraint on the mixing matrix for the neutrino mass
spectra (A) and (B).

The general case of mixing of four massive neutrinos is described by
$
\nu_{\alpha L}
=
\sum_{j=1}^4
U_{\alpha j} \ \nu_{jL} 
$,
where $U$ is the $4\times4$ unitary mixing matrix, $\alpha =
e,\mu,\tau,s$ denotes the three active neutrino flavors and the
sterile neutrino, respectively, and $j=1, \ldots, 4$ enumerates 
the neutrino mass eigenfields.
For definiteness, we will consider the spectrum of type I 
with a neutrino mass hierarchy
$ m_1 \ll m_2 \ll m_3 \ll m_4 $, but
the results that we will obtain in this case will 
apply to all spectra of class 1.

The probability of SBL $\nu_\mu\to\nu_e$
transitions is given by the two-neutrino-like formula
\cite{BGG96}
\begin{equation}
P_{\nu_\mu\to\nu_e}
=
P_{\bar\nu_\mu\to\bar\nu_e}
=
A_{\mu;e} \ \sin^2 \frac{\Delta m^2_{41} L}{4E}
\,,
\label{Pmue}
\end{equation}
where
$ \Delta m^2_{41} \equiv \Delta m^2_{\mathrm{LSND}} $,
$L$ is the distance between source and detector and $E$ is the
neutrino energy. We use the abbreviation 
$\Delta m^2_{kj} \equiv m_k^2 - m_j^2$.
The oscillation amplitude $A_{\mu;e}$ is given by
\begin{equation}\label{Amue}
A_{\mu;e}
=
4 \left( 1 - c_e \right) \left( 1 - c_\mu \right)
\end{equation}
with
\begin{equation}\label{calpha}
c_\alpha = \sum_{j=1}^3 |U_{\alpha j}|^2
\qquad
(\alpha = e , \mu)
\,.
\end{equation}
From the results of reactor and accelerator disappearance experiments
it follows that
\cite{BGG96}
\begin{equation}\label{ce}
c_\alpha \leq a^0_\alpha \quad \mbox{or} \quad c_\alpha \geq 1-a^0_\alpha
\end{equation}
with
$
a^0_\alpha
=
\frac{1}{2} \left( 1 - \sqrt{ 1 - B_{\alpha;\alpha}^0\, } \right)
$,
where
$B_{\alpha;\alpha}^0$
is the upper bound for the amplitude of $\nu_\alpha\to\nu_\alpha$
oscillations.
The exclusion plots obtained from the Bugey \cite{Bugey95}
and
CDHS \cite{CDHS84} and CCFR \cite{CCFR84} experiments
imply that
$ a_e^0 \lesssim 4 \times 10^{-2} $
for
$\Delta m^2_{\mathrm{LSND}} \gtrsim 0.1 \, \mathrm{eV}^2$
and
$ a_\mu^0 \lesssim 0.2 $
for
$\Delta m^2_{\mathrm{LSND}} \gtrsim 0.4 \, \mathrm{eV}^2$ \cite{a0}.
Below $\Delta m^2 \simeq 0.3$ eV$^2$, the survival
amplitude $B_{\mu;\mu}$ is
not restricted by experimental data, \textit{i.e.}, 
$B^0_{\mu;\mu}=1$.

The survival probability of solar $\nu_e$'s is bounded by 
$
P^\odot_{\nu_e\to\nu_e} \geq (1-c_e)^2
$ \cite{BGG96}.
Therefore,
to be in agreement with
the results of solar neutrino experiments
we conclude that from the two ranges of $c_e$ in Eq.(\ref{ce}) only
\begin{equation}\label{cee}
c_e \geq 1 - a^0_e
\end{equation}
is allowed.

We will address now the question of what information on the parameter
$c_\mu$ can be obtained from the asymmetry $A_\mu$ (\ref{A}).
As a first step we
derive an upper bound on the number of downward-going
$\mu$-like events $D$. The probability of
$\nu_\alpha\to\nu_\alpha$
and
$\bar\nu_\alpha\to\bar\nu_\alpha$
transitions of atmospheric neutrinos is given by
\begin{eqnarray}\label{survival}
&&
P_{\nu_\alpha\to\nu_\alpha}
=
P_{\bar\nu_\alpha\to\bar\nu_\alpha}
=
\nonumber
\\
&&
\left|\, \sum_{j=1,2} |U_{\alpha j}|^2 + 
|U_{\alpha 3}|^2 \exp \left( -i \frac{\Delta m^2_{31}L}{2E} \right)
\right|^2 + |U_{\alpha 4}|^4 \,,
\end{eqnarray}
where we have taken into account that 
$\Delta m^2_{41} \gg \Delta m^2_{31}$ 
and $\Delta m^2_{21}L/2E \ll 1$
($\Delta m^2_{21}$ is relevant for solar neutrinos).
Because of the small value of 
$\Delta m^2_{\mathrm{atm}} \equiv \Delta m^2_{31}$,
it is well fulfilled that downward-going neutrinos
do not oscillate with the atmospheric
mass-squared difference.\footnote{This is not completely true for
neutrino directions close to the horizon with
$\Delta m^2_\mathrm{atm} \gtrsim 3\times 10^{-3} \: \mathrm{eV}^2$. 
Taking into account the result of the CHOOZ experiment \cite{CHOOZ98}, we
have checked, however, that numerically this has a negligible impact
on the following discussion.}
Therefore, we obtain for the
survival probability of downward-going neutrinos
\begin{equation}\label{PD}
P^D_{\nu_\alpha\to\nu_\alpha} = c_\alpha^2 + (1-c_\alpha)^2 \,.
\end{equation}
Furthermore, conservation of probability and Eq.(\ref{cee})
allow to deduce the upper bound
\begin{equation}\label{bemu}
P^D_{\nu_e\to\nu_\mu} \leq 1 - P^D_{\nu_e\to\nu_e} =
2\, c_e (1-c_e) \leq 2\, a^0_e (1-a^0_e) \,.
\end{equation}
Note that all arguments hold for neutrinos and antineutrinos. Denoting the
number of muon (electron) neutrinos and antineutrinos 
produced in the atmosphere 
by $n_\mu$ ($n_e$), from Eqs.(\ref{PD}) and (\ref{bemu}) we have the
upper bound
\begin{equation}\label{Du}
D \leq n_\mu [ c_\mu^2 + (1 - c_\mu)^2 ] + 2\, n_e a^0_e (1-a^0_e) \,.
\end{equation}

Taking into account only the part of $D$ which is determined by the
$\stackrel{\scriptscriptstyle (-)}{\nu}_{\hskip-3pt \mu}$
survival probability, we immediately obtain the lower bound
\begin{equation}\label{Dl}
D \geq n_\mu [ c_\mu^2 + (1 - c_\mu)^2 ] \,.
\end{equation}
Considering only $|U_{\mu 4}|^4$ in Eq.(\ref{survival}),
we readily arrive at a lower bound on $U$ as well:
\begin{equation}\label{Ul}
U \geq n_\mu (1-c_\mu)^2 \,.
\end{equation}
This inequality is analogous to the above inequality for the survival
of solar neutrinos and is valid also with matter effects in the earth.

\begin{figure}[t]
\epsfig{file=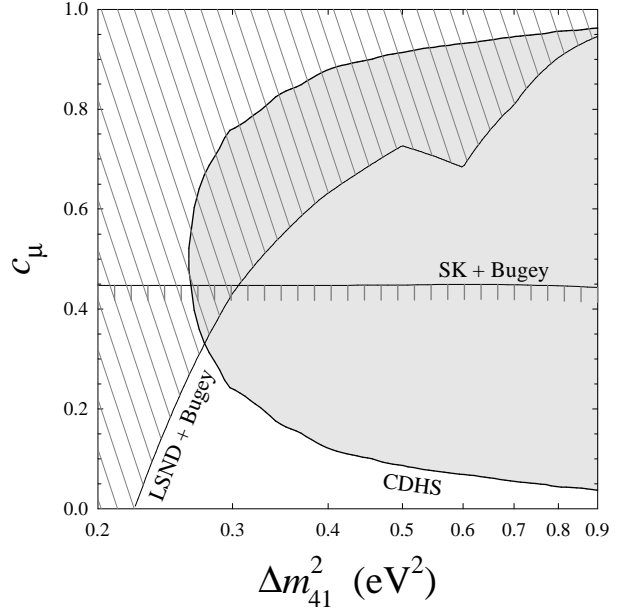,width=0.95\linewidth}
\caption{ \label{cmu}
Regions in the $\Delta{m}^2_{41}$--$c_\mu$
plane disfavored by the results of the CDHS,
LSND, Super-Kamiokande and Bugey experiments in the case of the
spectra of class 1. The shaded region is excluded by the inequalities
(\ref{cm}) and the hatched region by the bound (\ref{LSND}). The nearly
horizontal curve labelled SK + Bugey represents the lower bound
(\ref{ineq}) derived from the Super-Kamiokande up--down asymmetry. Since
this bound lies above the white region allowed by inequalities
(\ref{cm}) and (\ref{LSND}), the spectra of class 1 are disfavored by the data.}
\end{figure}

Now we can assemble the inequalities (\ref{Du}), (\ref{Dl}) and
(\ref{Ul}) and it follows the main result of this work
\begin{equation}\label{ineq}
-A_\mu \leq \frac{c_\mu^2 + 2\, a^0_e (1-a^0_e) /r}{c_\mu^2 + 2(1-c_\mu)^2} \,,
\end{equation}
where we have defined $r \equiv n_\mu/n_e$. For the numerical
evaluation of Eq.(\ref{ineq}) we use $-A_\mu \geq 0.254$ at 90\% CL,
the 90\% CL bound $a^0_e$ from the result of the Bugey experiment and $r = 2.8$
read off from Fig.~3 in Ref.~\cite{SK-atm-98} of the Super-Kamiokande
Collaboration. As a result we get 
\begin{equation}
c_\mu \geq a_{\mathrm{SK}} \simeq 0.45
\,, 
\end{equation}
as can be
seen from the horizontal line in Fig.~\ref{cmu}. Note that the
dependence of this lower bound on 
$\Delta m^2_{\mathrm{LSND}} \equiv \Delta m^2_{41}$ is almost
negligible due to the smallness of the second term in the
numerator on the right-hand side of Eq.(\ref{ineq}).
Consequently, also the exact value of $r$ is not important numerically.

In Fig.~\ref{cmu} we have also depicted the bounds
\begin{equation}\label{cm}
c_\mu \leq a^0_\mu \quad \mbox{and} \quad c_\mu \geq 1-a^0_\mu
\end{equation}
that were obtained from the exclusion plot of
the CDHS $\nu_\mu$ disappearance experiment. 
For $\Delta m^2_{\mathrm{LSND}} \simeq 0.24$ eV$^2$ these two
bounds meet at $c_\mu = 0.5$. Below 0.24 eV$^2$ there are no
restrictions on $c_\mu$ from SBL experiments.

Finally, we take into account the result of the LSND experiment, from
which information on the SBL $\bar \nu_\mu \to \bar \nu_e$ 
transition amplitude $A_{\mu;e}$ (\ref{Amue}) is obtained.
Using Eq.(\ref{cee}) and the lower bound
$A^\mathrm{min}_{\mu;e}$, which can be inferred from the 
region allowed by LSND,
we derive the further bound on $c_\mu$ \cite{BGG98-Pune}
\begin{equation}\label{LSND}
c_\mu \leq a_\mathrm{LSND} \equiv 1 - A^\mathrm{min}_{\mu;e}/4a^0_e \,.
\end{equation}
This bound is represented by the curve in Fig.~\ref{cmu} labelled
LSND + Bugey. 

Fig.~\ref{cmu} clearly shows that a four-neutrino mass hierarchy is
strongly disfavored because no allowed region for $c_\mu$ is
left in this plot. A four-neutrino mass hierarchy is also strongly
disfavored for $\Delta m^2_{\mathrm{LSND}} \gtrsim 0.4$
eV$^2$ as was shown in Ref.~\cite{BGG96}. We want to stress that all bounds are
derived from 90\% CL plots and that the bound (\ref{LSND}) is quite
sensitive to the actual values of $A^\mathrm{min}_{\mu;e}$ and 
$a^0_e$. This has to be kept in mind in judging the result derived
here. As was noticed before \cite{BGG96}, 
the procedure discussed here applies to
all four-neutrino mass spectra of class 1 
where a group of three neutrino
masses is close together and separated from the fourth neutrino
mass by a gap needed to explain the result of the LSND
experiment. The reason is that all arguments presented here
remain unchanged if one defines $c_\alpha$ (\ref{Amue}) by a
summation over the indices of the three close masses for each of the mass
spectra of class 1 (see Fig.~\ref{4spectra}), \textit{i.e.},
$j=1,2,3$ for the spectra I and II and $j=2,3,4$ for the spectra
III and IV.

To give an intuitive understanding that the data disfavor all spectra
of class 1 we note that $c_\mu$ cannot be too close to 1 in order to
explain the non-zero LSND $\bar\nu_\mu\to\bar\nu_e$ oscillation
amplitude (\ref{Amue}). On the other hand, if $c_\mu$ is too close to
zero, the atmospheric $\nu_\mu$ oscillations are suppressed 
(see Eq.(\ref{survival}), taking into account that 
$|U_{\mu 4}|^2 = 1-c_\mu$). For 
$\Delta m^2_\mathrm{LSND} \lesssim 0.3 \: \mathrm{eV}^2$ these two
requirements contradict each other. For
$\Delta m^2_\mathrm{LSND} \gtrsim 0.3 \: \mathrm{eV}^2$ they are in
contradiction to the results of the CDHS and CCFR $\nu_\mu$
disappearance experiments requiring $c_\mu$ to be either close to zero
or 1 (see Eq.(\ref{ce})).

According to the previous discussion, only the mass spectra of
class 2 remain. They can be characterized in the following way:
\begin{equation} \label{(A)}
\mbox{(A)}
\qquad
\underbrace{
\overbrace{m_1 < m_2}^{\mathrm{atm}}
\ll
\overbrace{m_3 < m_4}^{\mathrm{solar}}
}_{\mathrm{LSND}}
\end{equation}
and
\begin{equation} \label{(B)}
\mbox{(B)}
\qquad
\underbrace{
\overbrace{m_1 < m_2}^{\mathrm{solar}}
\ll
\overbrace{m_3 < m_4}^{\mathrm{atm}}
}_{\mathrm{LSND}}
\;.
\end{equation}

Let us now discuss which impact the up--down asymmetry $A_\mu$ has
on these mass schemes. We consider first scheme (A) and go through the 
same steps as in the case of the mass
hierarchy. Now we define
\begin{equation}\label{caA}
c_\alpha = \sum_{j=1,2} |U_{\alpha j}|^2 \,.
\end{equation}
Then the results of reactor experiments and the  
energy-dependent suppression of the solar neutrino flux leads to
\begin{equation}
c_e \leq a^0_e \,.
\end{equation}
Repeating the derivation of Eq.(\ref{ineq}) with $c_\alpha$ as
defined in Eq.(\ref{caA}), it is easily seen that the inequality
(\ref{ineq}) holds also for scheme (A). On the other hand, the bound that
takes into account the LSND result now has the form
\begin{equation}\label{LSNDA}
c_\mu \geq A^\mathrm{min}_{\mu;e}/4a^0_e \,.
\end{equation}
The corresponding curve in the $\Delta m^2_{41}$--$c_\mu$ plane is given
by a reflection of the
curve labelled LSND + Bugey in Fig.~\ref{cmu} at the horizontal line
$c_\mu = 0.5$. Therefore, in the case of scheme (A) the allowed region
of $c_\mu$ is determined by the bound (\ref{LSNDA}) and by $c_\mu \geq
1-a^0_\mu$. This region is allowed and not restricted by 
$c_\mu \gtrsim 0.45$ obtained from the Super-Kamiokande up--down
asymmetry.

A discussion of scheme (B) with $c_e \geq 1-a^0_e$ leads to the bound
(\ref{ineq}) with $c_\mu$ replaced by $1-c_\mu$ in this formula and
to Eq.(\ref{LSND}). Therefore, the bounds for scheme (B) are obtained
from those of scheme (A) by a reflection of the curves at the line
$c_\mu = 0.5$. In summary, the white area in Fig.~\ref{cmu}
represents the allowed region for $1-c_\mu$ in scheme (A) and for
$c_\mu$ in scheme (B).

In this paper we have shown that the existing neutrino oscillation data
allow to draw definite conclusions 
about the nature of the possible four-neutrino
mass spectra. We have demonstrated that the spectra (I) -- (IV) in
Fig.~\ref{4spectra}, including the hierarchical one, 
are all disfavored by the data in the whole range
$0.2 \: \mathrm{eV}^2 \lesssim \Delta m^2_\mathrm{LSND} \lesssim 2 \:
\mathrm{eV}^2$ of the mass-squared difference determined by LSND and
other SBL neutrino oscillation experiments. With the Super-Kamiokande
result on the atmospheric up--down asymmetry it has been also possible to
investigate the region 
$\Delta m^2_\mathrm{LSND} \lesssim 0.3 \: \mathrm{eV}^2$
which was not explored in previous publications. The only
four-neutrino mass spectra that can accommodate all the existing neutrino
oscillation data are the spectra (A) and (B) in Fig.~\ref{4spectra} in
which two pairs of close masses are separated by the LSND
mass gap. The analysis introduced in this paper enables us in addition
to obtain information on the mixing matrix $U$ via a rather stringent
bound on the quantity $c_\mu$ (\ref{caA}) for the allowed schemes (A) and (B).

\acknowledgments
S.M.B. would like to thank the Institute for Theoretical Physics of
the University of Vienna for its hospitality.

\end{document}